\documentclass[runningheads]{llncs}
\usepackage{graphicx}

\begin{document}

\title{Registration of Histopathogy Images Using Structural Information From Fine Grained Feature Maps }

\author{Dwarikanath Mahapatra
}

\authorrunning{Mahapatra}

\institute{Inception Institute of Artificial Intelligence, Abu Dhabi, UAE}
%




%
\maketitle              
%
\begin{abstract}
  Registration is an important part of many clinical workflows and factually, including information of structures of interest improves registration performance. We propose a novel approach of combining segmentation information in a registration framework  using self supervised segmentation feature  maps extracted using a pre-trained segmentation network followed by clustering. Using self supervised feature maps enables us to use segmentation information despite the unavailability of manual segmentations. Experimental results show our approach effectively  replaces manual segmentation maps and demonstrate the possibility of obtaining state of the art registration performance in real world cases where manual segmentation maps are unavailable.
   
  \keywords{Fine grained segmentation  \and Registration \and Histopathology.}
   
\end{abstract}
%


\section{Introduction}
\label{sec:intro}

An important part of a digital pathology image analysis workflow is the  visual comparison of successive tissue sections enabling pathologists to evaluate histology and expression of multiple markers in a single area \cite{Rep_2}. It has the potential to improve diagnostic accuracy and requires aligning all images to a common frame. Due to tissue processing and other procedures certain sections may undergo elastic deformations leading to shape changes across slices. Thus it is essential to have a reliable registration algorithm that can address challenges such as: (i) registration despite  differences in tissue appearance (a consequence of different dyes and sample preparation); (ii) computing time due to large microscopy images; (iii) correcting  complex elastic deformations, and (iv) difficulty of global unique features due to repetitive texture \cite{ANHIR}.

A comprehensive review on conventional image registration methods can be found in \cite{RegRev,Frontiers2020,Mahapatra_PR2020,ZGe_MTA2019,Behzad_PR2020,Mahapatra_CVIU2019,Mahapatra_CMIG2019}. Although widely used, these methods are time consuming due to: 1) iterative optimization techniques; and 2) extensive parameter tuning. Deep learning (DL) methods can potentially overcome  these limitations by using trained models to output registered images and deformation fields in much less time. 
In recent DL based image registration works, Sokooti et. al. \cite{RegNet,Mahapatra_LME_PR2017,Zilly_CMIG_2016,Mahapatra_SSLAL_CD_CMPB,Mahapatra_SSLAL_Pro_JMI,Mahapatra_LME_CVIU} propose RegNet using CNNs trained on simulated deformations to register a pair of unimodal images.
 Vos et. al. \cite{Vos_DIR,LiTMI_2015,MahapatraJDI_Cardiac_FSL,Mahapatra_JSTSP2014,MahapatraTIP_RF2014,MahapatraTBME_Pro2014} propose the deformable image registration network (DIRNet) which outputs a transformed image non-iteratively without having to train on known registration transformations. 
%
Rohe et. al. \cite{RoheMICCAI2017,MahapatraTMI_CD2013,MahapatraJDICD2013,MahapatraJDIMutCont2013,MahapatraJDIGCSP2013,MahapatraJDIJSGR2013} propose SVF-Net that uses deformations obtained by registering previously segmented regions of interest (ROIs). 
The above methods are limited by the need of spatially corresponding patches (\cite{RegNet,Vos_DIR,MahapatraJDISkull2012,MahapatraTIP2012,MahapatraTBME2011,MahapatraEURASIP2010}) or being too dependent on training data.
Balakrishnan et. al. \cite{BalaCVPR18,Mahapatra_CVPR2020,Kuanar_ICIP19,Bozorgtabar_ICCV19,Xing_MICCAI19,Mahapatra_ISBI19} learn a parameterized registration function from a collection of training volumes, and in \cite{VMorphTMI,MahapatraAL_MICCAI18,Mahapatra_MLMI18,Sedai_OMIA18,Sedai_MLMI18,MahapatraGAN_ISBI18} they improve on their method by adding segmentation information. 

 Hu et al. in \cite{HuMic19_CondSeg,Sedai_MICCAI17,Mahapatra_MICCAI17,Roy_ISBI17,Roy_DICTA16,Tennakoon_OMIA16} propose a paradigm where registration is posed as a problem of segmenting corresponding regions of interest (ROIs) across images. 
 Lee et al. in \cite{LeeMic19_STN,Sedai_OMIA16,Mahapatra_MLMI16,Sedai_EMBC16,Mahapatra_EMBC16,Mahapatra_MLMI15_Optic} register anatomical structures by mapping input images to spatial transformation parameters using Image-and-Spatial Transformer Networks (ISTNs).
Liu et al. \cite{LiuMic19_Brain3d,Mahapatra_MLMI15_Prostate,Mahapatra_OMIA15,MahapatraISBI15_Optic,MahapatraISBI15_JSGR,MahapatraISBI15_CD} propose feature-level probabilistic model to regularize CNN hidden layers for 3D brain image alignment. Hu et al. \cite{HuMic19_Dual,KuangAMM14,Mahapatra_ABD2014,Schuffler_ABD2014,MahapatraISBI_CD2014,MahapatraMICCAI_CD2013,Schuffler_ABD2013} use a two-stream 3D encoder-decoder network and pyramid registration module for registration.

It is a well established fact that registration and  segmentation are inter-related and mutually complementary. 
For example, labeled atlas images are used via image registration for segmentation while segmentation maps provide extra information to aid in image registration and result evaluation. Methods combining registration and segmentation have been proposed using active-contours \cite{ZXuMic19_11,MahapatraProISBI13,MahapatraRVISBI13,MahapatraWssISBI13,MahapatraCDFssISBI13,MahapatraCDSPIE13,MahapatraABD12}, Bayesian methods \cite{ZXuMic19_7,MahapatraMLMI12,MahapatraSTACOM12,VosEMBC,MahapatraGRSPIE12,MahapatraMiccaiIAHBD11,MahapatraMiccai11} or Markov random field formulations \cite{ZXuMic19_6,MahapatraMiccai10,MahapatraICIP10,MahapatraICDIP10a,MahapatraICDIP10b,MahapatraMiccai08,MahapatraISBI08}. Recent deep learning based approaches have used GANs \cite{Mahapatra_MLMI2018,MahapatraICME08,MahapatraICBME08_Retrieve,MahapatraICBME08_Sal,MahapatraSPIE08,MahapatraICIT06} and a Deep Atlas \cite{ZXuMiccai19,CVPR2020_Ar,sZoom_Ar,CVIU_Ar,AMD_OCT,GANReg1_Ar,PGAN_Ar} for joint registration and segmentation.

\subsection{Contributions}

While these methods show the advantages of integrating segmentation information with registration, they are dependent upon the availability of annotated segmentation maps during training. We propose a method to integrate structural  information from self-supervised segmentation maps for registering histopathological images. Our approach does not require manual segmentation maps during training and test times.
 Self supervised approaches generate detailed segmentation feature maps of an image pair that provide fine grained information about the structures of interest. The structural information is included in the registration framework by means of a loss function that captures the structural similarity between the image pair being registered.  Inclusion of the segmentation information contributes significantly to improved registration by providing the additional structural information that is so important for medical images. These maps are generated for the reference and floating image using a pre-trained semantic segmentation network and can be used for new images during test time. 

\section{Method}
\label{sec:method}

\begin{figure}[t]
 \centering
\begin{tabular}{c}
\includegraphics[height=3.5cm, width=10.1cm]{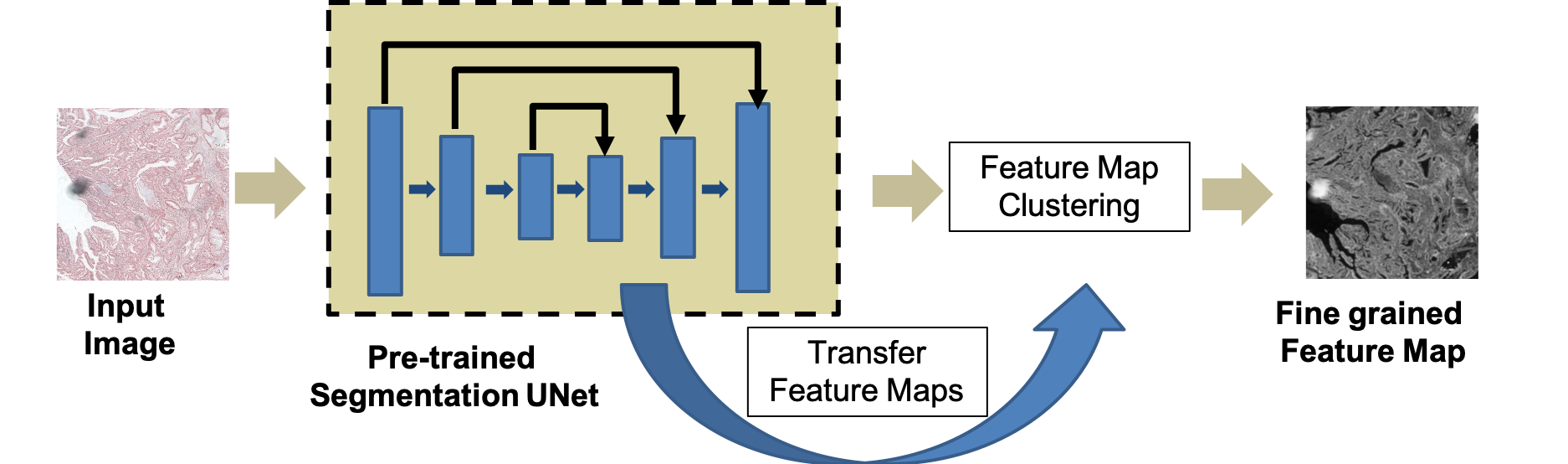}  \\
(a) Generation of Fine grained feature maps \\
\includegraphics[height=3.5cm, width=10.1cm]{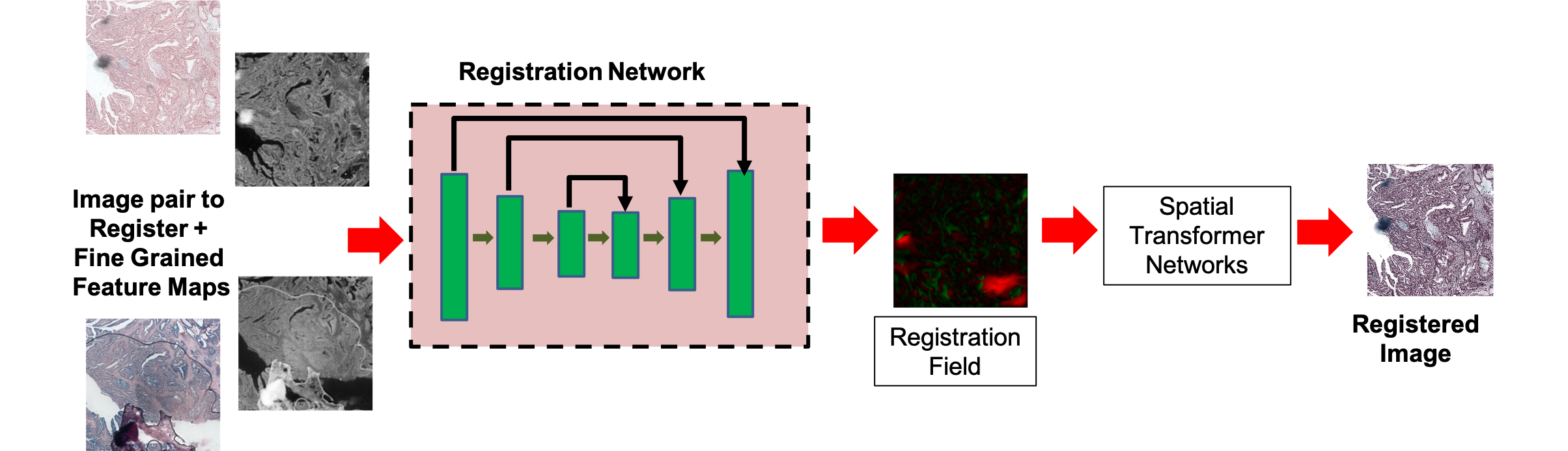}  \\
(b) Depiction of registration workflow \\
\end{tabular}
\caption{Depiction of proposed workflow for generating fine grained feature maps and registration output.}
\label{fig:overview}
\end{figure}

Figure~\ref{fig:overview} shows the workflow of our proposed approach. The reference image ($I_{R}$) and floating image ($I_{F}$) are passed through the pre-trained segmentation network to generate segmentation maps $I_{R}^M$ and $I_F^{M}$. The maps and original images are input to a UNet architecture based network that outputs a registration field used by a spatial transformer network (STN) to output the transformed image. 

\subsection{Self Supervised Segmentation Feature Maps}
\label{met:sseg}

Self-supervised learning is a variant of unsupervised learning methods,
where a model learns to predict a set of labels that can be automatically created from the input data. A general approach is to train a CNN to perform an auxiliary task such as visual localization \cite{Larsson,Haze_Ar,Xr_Ar,RegGan_Ar,ISR_Ar,LME_Ar,Misc}, predicting missing image parts \cite{Larsson60,Health_p,Pat2,Pat3,Pat4,Pat5,Pat6}, and cardiac MR image segmentation \cite{BaiMiccai19,Pat7,Pat8,Pat9,Pat10}. Defining an auxiliary task is dataset dependent and the same task/model may not be applicable when we aim to generalize the model to new dataset. Different from previous methods we generate self supervised segmentation maps by first extracting feature maps from a pre-trained segmentation network and then apply k-means clustering to generate a very detailed fine grained feature map.

We use a Fine-Grained Segmentation Network (FGSN) with 
a backbone UNet \cite{UNet,Pat11,Pat12,Pat13,Pat14,Pat15}, that has been pre-trained on the Glas dataset \cite{GlasReview} for segmenting histopathological images. The architecture, shown in Figure~\ref{fig:overview} (a) has $3$ convolution blocks each in the contracting (Encoder) and expanding (Decoder) path (shown as blue blocks). Each convolution block has $3$ convolution layers of $3\times3$  filters, with ReLu activation and batch normalization. Each block is followed by $2\times2$ max pooling.
Given the images $I_{R},I_{F}$, we  generate segmentation maps $I_{R}^{M},I_{F}^{M}$. 
First  we concatenate the maps of corresponding layers of the encoder and decoder part, upsample and fuse them to get feature maps. This allows us to integrate information from multiple scales. 

After obtaining the feature maps we apply $k$-means clustering and the cluster assignments  are used as labels. We can change the number of classes that the FGSN outputs without having to create new annotations. Higher the value of $k$ denser are the segmentation maps. 
The initial feature maps are first clustered with a high value of $k=\frac{3N}{4}$ where $N$ is the number of pixels, and reduce it as $k_{t+1}=0.9k_t$, $t$ being the iteration index. The optimum cluster number is determined using the Gap statistic method of \cite{Gap_stat} having the following steps: 

\begin{enumerate}
    \item Cluster the observed data by varying the number of clusters $k$ and compute the corresponding total within intra-cluster variation $W_k$.
    \item Generate $B$ reference data sets with a random uniform distribution. Cluster each of these reference data sets with varying number of clusters, and compute the corresponding total within intra-cluster variation $W_{kb}$.
    \item Compute the estimated gap statistic as the deviation of the observed $W_k$ value from its expected value $W_{kb}$ under the null hypothesis: $Gap(k)=\frac{1}{B} \sum_{b=1}^{B} \log (W^{*}_{kb}) - \log (W_k)$. Also compute the standard deviation.
    \item Choose the number of clusters as the smallest value of $k$ such that the gap statistic is within one standard deviation of the gap at k+1, i.e., $Gap(k)\geq Gap(k+1) - s_{k+1}$.
\end{enumerate}


\subsection{Registration using Segmentation Maps}
\label{met:reg}

Our primary contribution is the integration of segmentation features, generated in a self-supervised manner, into a deep learning based registration framework. Hence our approach is framework agnostic and can be used with various architectures. However, we choose to demonstrate our method's effectiveness using the VoxelMorph architecture \cite{VMorphTMI} because of two reasons: 1) In \cite{VMorphTMI} the authors apply it for brain atlas registration on public datasets having manual segmentation maps. As a result they use a dice loss between manual segmentation maps in their training loss function. We aim to investigate if we can effectively replace the manual maps with our self supervised segmentation maps for the same image dataset. 2) secondly, since VoxelMorph has proven to be a popular baseline method for many works we investigate the improvements brought about by adding an additional segmentation loss from self supervised segmentation maps.

Figure~\ref{fig:overview} (b) shows the registration network architecture. 
Given reference and floating images  $I_R$, $I_F$, we assume they have been affinely aligned, and only non-linear displacements exist between them. 
A Encoder-Decoder architecture is used for computing the deformation field $\varphi$ which is used to transform $I_F$ and obtain the registered image $I_{Reg}=\varphi\circ I_F$ through spatial transformer networks (STNs) \cite{STN}. In an ideal registration $I_R$ and $I_{Reg}$ should match. 
%
 Similar to the UNet architecture each block represents set of $3$ 2D convolution layers of kernel size $3$, stride $2$, followed by a LeakyReLU layer with parameter $0.2$.
 The optimal parameter values are learned 
 by minimizing differences between $I_{Reg}$ and $I_R$ . In order
to use standard gradient-based methods, we construct a differentiable operation based on STN \cite{STN} 




\subsubsection{Loss Functions:}

We use  two loss functions: an unsupervised loss $L_{us}$ and an auxiliary loss
$L_a$ that leverages anatomical segmentations at training time.
The unsupervised loss consists of two components: $L_{sim}$ that penalizes
differences in appearance, and $L_{smooth}$ that penalizes local
spatial variations in $\varphi$:
\begin{equation}
    L_{us} (I_R, I_F, \varphi) = L_{sim} (I_R, I_{Reg}) + \lambda_1L_{smooth} (\varphi) + \lambda_2L_{seg}(M_R,M_F,M_{Reg}),
    \label{eq:loss}
\end{equation}
where $\lambda_1=0.95, \lambda_2=1.05$ are regularization parameters. $L_{sim}$ is a combination of mean squared pixel intensity difference, and local cross correlation between $I_R, I_{Reg}$. $L_{seg}$ is the segmentation loss function and is defined as 
\begin{equation}
L_{seg}(M_R,M_{Reg})=MSE(M_R,M_{Reg})
\label{eq:segloss}
\end{equation}
which is the pixel wise difference between the segmentation maps of the reference and registered image. $M_{Reg}=\varphi\circ M_F$ is obtained by applying the deformation field $\varphi$ to the segmentation map of the floating image being registered.  
A smooth displacement field $\varphi$ is obtained using a diffusion regularizer on the spatial gradients of displacement u:
\begin{equation}
   L_{smooth}(\varphi) = \sum_{p\in \Omega} \left\|\nabla u(p) \right\|^{2}
\end{equation}



\section{Experimental Results}
\label{sec:expt}

\subsection{Implementation and Dataset Details}

Our method was implemented in TensorFlow. We use Adam \cite{Adam} with $\beta_1=0.93$ and batch normalization.  The network was trained with $10^{5}$ update iterations at learning rate $10^{-3}$. Training and test was performed on a NVIDIA Titan X GPU with $12$ GB RAM.
We test our method on two datasets: 1) the publicly available ANHIR challenge dataset \cite{ANHIR} used to evaluate automatic nonlinear image registration of 2D microscopy images of histopathology tissue samples stained with different dyes; and 2) a subset of the brain images used in \cite{VMorphTMI}. Their descriptions are given below. 

\textbf{ANHIR Dataset}: 
The dataset consists of $481$ image pairs and is split into
$230$ training and $251$ evaluation pairs \cite{ANHIR}. For the
training image pairs, both source and target landmarks are
available, while for evaluation image pairs only source
landmarks are available. 
The landmarks were annotated by $9$ experts and there are on an average $86$ landmarks per
image. The average error between the same landmarks chosen
by two annotators is $0.05\%$ of the image size which can be
used as the indicator of the best possible results to achieve
by the registration methods, below which the results become
indistinguishable \cite{Rep7}.

The images are divided into $8$ classes containing: (i) lesion
tissue, (ii) lung lobes, (iii) mammary glands, (iv) the colon
adenocarcinomas, (v) mice kidney tissues, (vi) gastric mucosa
and adenocarcinomas tissues, (vii) breast tissues, (viii) human
kidney tissues. All the tissues were acquired in
different acquisition settings making the dataset more diverse.
The challenge was performed on medium
size images which were approximately $25\%$ of the original
image scale. The approximate image size after the resampling
varied from $8k$ to $16k$ pixels (in one dimension).

\textbf{Brain Image Dataset:}
%
%
We used the $800$ images of the ADNI-1 dataset \cite{Bala33} consisting of $200$ controls, $400$ MCI and $200$ Alzheimer's Disease patients. The MRI protocol for ADNI1 focused on consistent longitudinal structural imaging on $1.5T$ scanners using $T1$ and dual echo $T2-$weighted sequences. 
All scans were resampled to $256\times256\times256$ with $1$mm isotropic voxels. Pre-processing includes affine registration and brain extraction using FreeSurfer \cite{Bala17}, and cropping the resulting images to $160 \times 192 \times 224$. 
The dataset is split into $560$, $120$, and $120$ volumes for training, validation, and testing. We simulate elastic deformations to generate the floating images  using which we evaluate registration performance. Registration performance was calculated for $2$D slices. 



\subsection{ANHIR Registration Results}

Registration performance is evaluated using a normalized version of target registration error (TRE) and is defined as:
\begin{equation}
    rTRE=\frac{TRE}{\sqrt{w^{2}+h^{2}}}
\end{equation}
where $TRE$ is the target registration calculated as the Euclidean distance between corresponding landmarks in the two images and $w,h$ are the image height and width respectively.

Table~\ref{tab:regres} summarizes the performance of $SR-Net$ (our prposed Segmentation based Registration NETwork) and the top $2$ methods  on different tissue types. The median rTRE for all tissues and each individual tissue type is reported with the numbers taken from \cite{ANHIR}.

We perform a set of ablation experiments where we exclude the segmentation loss of Eqn.\ref{eq:segloss} (denoted as $SR-Net_{wL_{seg}}$), using either MSE or cross correlation in $L_{sim}$ of Eqn.\ref{eq:loss} (denoted, respectively, $SR-Net_{MSE}$, $SR-Net_{MSE}$). Since our registration framework is similar to VoxelMorph, $SR-Net_{wL_{seg}}$ is equivalent to VoxelMorph without the segmentation loss.  The results show that SR-Net outperforms the top ranked methods for the challenge. Moreover, the advantages of using segmentation is also obvious by the fact that the rTRE values for SR-Net$_{wL_{Seg}}$
are significantly higher than SR-Net ($p=0.003$ from a paired Wilcoxon test). Ablation experiments also quantify the 
contribution of MSE and CC in the registration framework.

\begin{figure}[h]
 \centering
\begin{tabular}{ccccc}
\includegraphics[height=2.6cm, width=2.3cm]{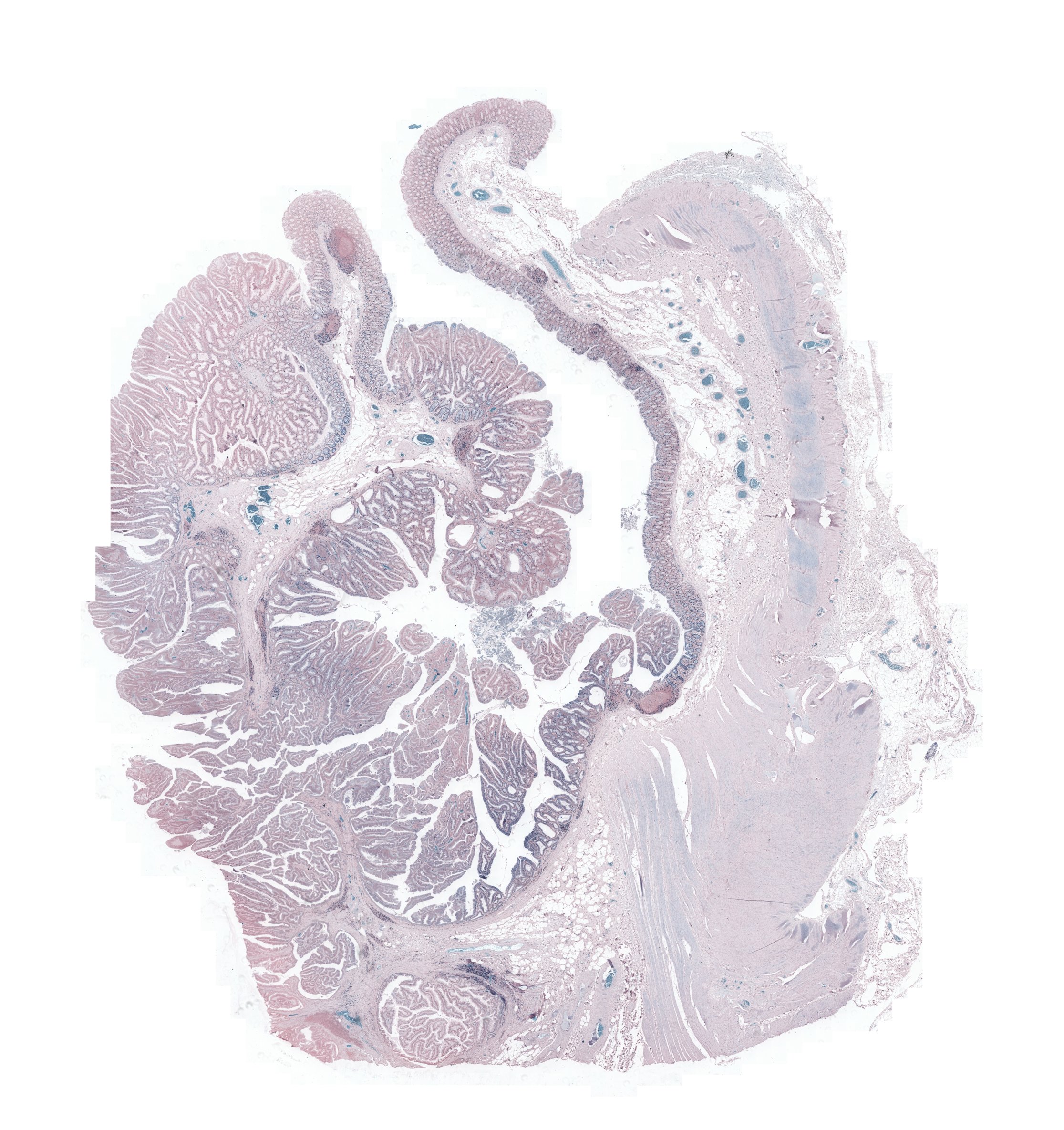} & 
\includegraphics[height=2.6cm, width=2.3cm]{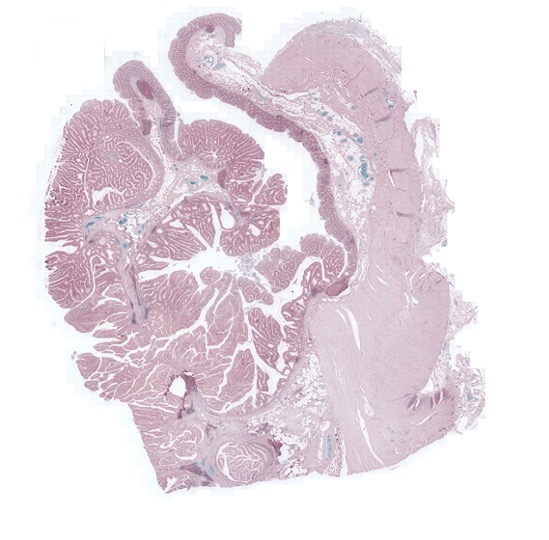} & 
\includegraphics[height=2.6cm, width=2.3cm]{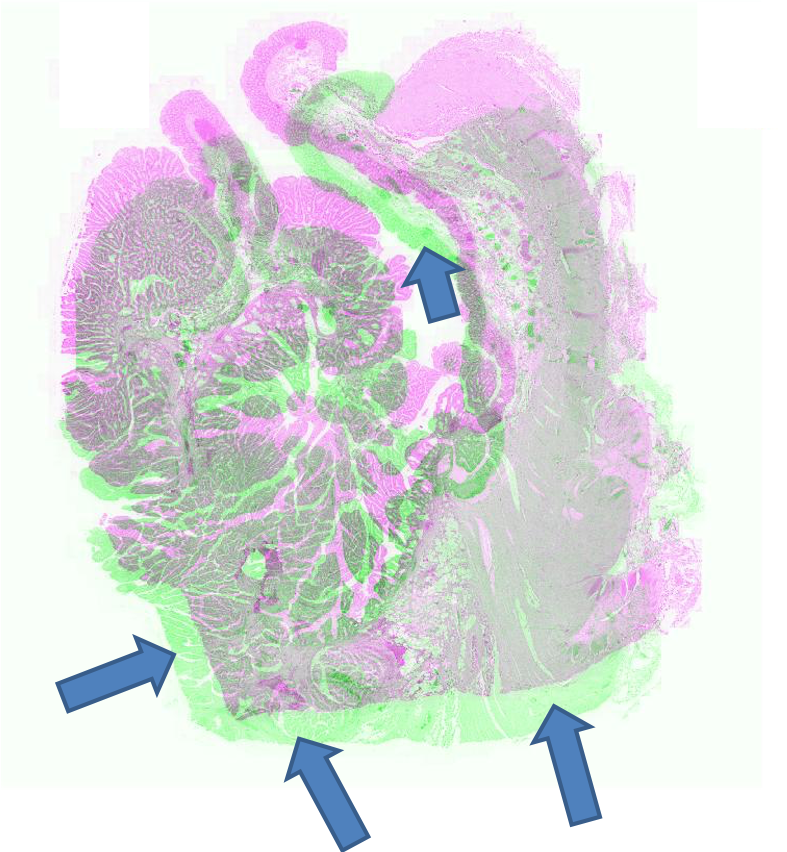} & 
\includegraphics[height=2.6cm, width=2.3cm]{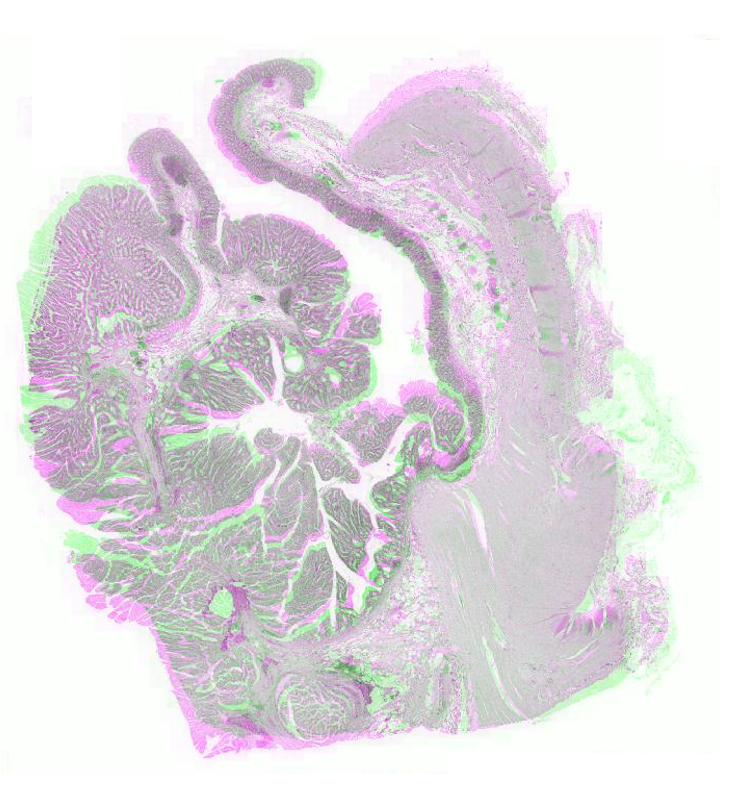} & 
\includegraphics[height=2.6cm, width=2.3cm]{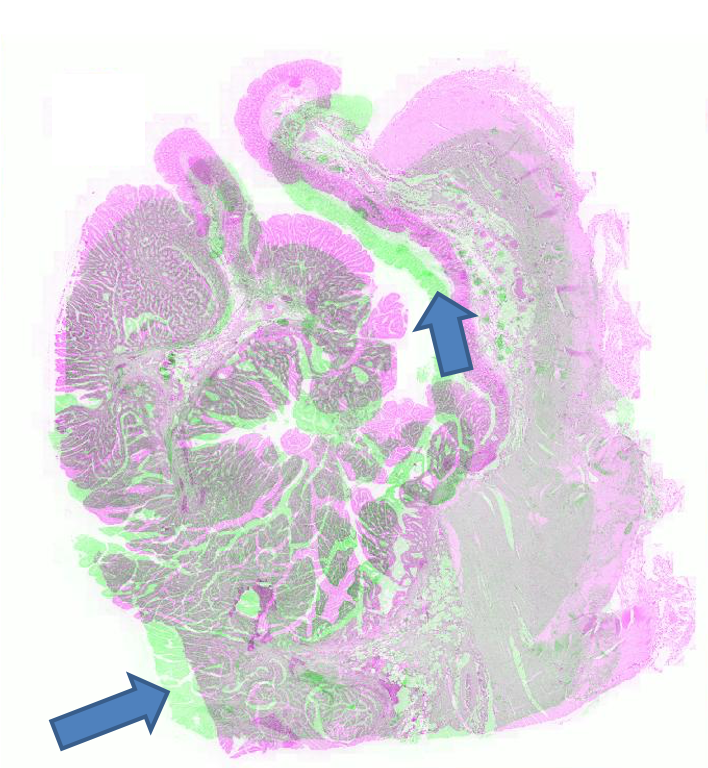} \\ 
(a) & (b) & (c) & (d) & (e)\\ 
\end{tabular}
\caption{Pathology image registration result: (a) $I_R$; (b) $I_F$; Misalignment of images: (c) before registration; after registration using (d) SR-Net; (e) SR-Net$_{wL_{Seg}}$}
\label{fig:pathres}
\end{figure}

Figure~\ref{fig:pathres} shows the registration results for pathology images where we show the reference and floating images alongwith the misalignment images before registration and after registration using SR-Net and SR-Net$_{wL_{Seg}}$. The  misalignment is greatly reduced after registration using SR-Net while in the case of SR-Net$_{wL_{Seg}}$ there is still some resulting misalignment. This error can have significant consequences in the final diagnosis workflow. Hence the advantages of self-supervised segmentation maps are quite clear. 


\begin{table}[t]
 \begin{center}
\begin{tabular}{|c|c|c|c|c|c|c|}
\hline 
{} & {$SR-Net$} & {Rank~1}  & {Rank~2} & {SR-Net$_{wSeg}$} & {SR-Net$_{MSE}$} & {SR-Net$_{CC}$}  \\ \hline
{All} & {0.00099} & {0.00106}  & {0.00183} & {0.00202} & {0.00196} & {0.00193}  \\ \hline
{COADs} & {0.00123} & {0.00155}  & {0.00198} & {0.00231} & {0.00214} & {0.00217}  \\ \hline
{Breast tissue} & {0.00198} & {0.00222}  & {0.00248} & {0.00293} & {0.00259} & {0.00263}  \\ \hline
{human kidney} & {0.00231} & {0.00252}  & {0.00259} & {0.00284} & {0.00273} & {0.00274}  \\ \hline
{Gastric} & {0.00091} & {0.00122}  & {0.00061} & {.00154} & {0.00139} & {0.00137}  \\ \hline
{Lung lobes} & {0.00011} & {0.00008}  & {0.00138} & {0.00167} & {0.00151} & {0.00157}  \\ \hline
{Lung lesion} & {0.00010} & {0.00012}  & {0.00534} & {.00591} & {0.00552} & {0.00567}  \\ \hline
{Mice kidneys} & {0.00098} & {0.00104}  & {0.00137} & {0.00171} & {0.00152} & {0.00160}  \\ \hline
{Mammary gland} & {0.00017} & {0.00011}  & {0.00266} & {0.00281} & {0.00273} & {0.00269}  \\ \hline
\end{tabular}
\caption{Registration errors for different methods on the ANHIR dataset including ablation experiments. Values for $Rank~1,Rank~2$ are taken from \cite{ANHIR}}
\label{tab:regres}
\end{center}
\end{table}

\subsection{Brain Image Registration}

We apply $SR-Net$ to the brain image dataset. $SR-Net$'s difference w.r.t VoxelMorph is the use of self-supervised segmentation maps instead of manually annotated maps. The result is summarized in Table~\ref{tab:brainReg}. SR-Net comes close to Voxel Morph's performance and there is no statistically significant difference between the results ($p=0.031$). The numbers clearly show the significant improvement brought about by self supervised segmentation maps, to the extent that it performs at par when manual segmentation maps are used. VoxelMorph results can be considered the best possible performance under the given conditions and we are able to reach that with no manual segmentation maps.


\begin{table}[t]
\begin{tabular}{|c|c|c|c|c|c|c|}
\hline
{} & {Before} & \multicolumn {5}{|c|}{After Registration}  \\ \cline{3-7} 
{} & {Registration} & {SR-Net}  & {SR-Net$_{wL_{Seg}}$} & {DIRNet} & {FlowNet} & {VoxelMorph}   \\ \hline
%
{DM($\%$)} & {67.2} & {79.2}  & {72.6} & {73.0} & {72.8} & {79.5}\\ \hline
{HD$_{95}$(mm)} & {14.5} & {12.1}  & {13.7} & {13.4} & {13.5} & {11.9} \\ \hline
{Time(s)} & {} & {0.5} & {0.4} & {0.6} & {0.5} & {0.5} \\ \hline
\end{tabular}
\caption{Registration results for different methods on brain images. }
\label{tab:brainReg}
\end{table}



\begin{figure}[t]
\begin{tabular}{ccccc}
\includegraphics[height=2.4cm,width=2.3cm]{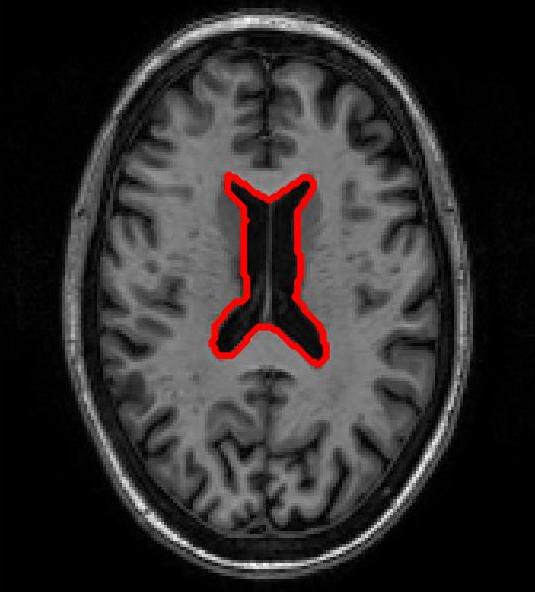} &
\includegraphics[height=2.4cm,width=2.3cm]{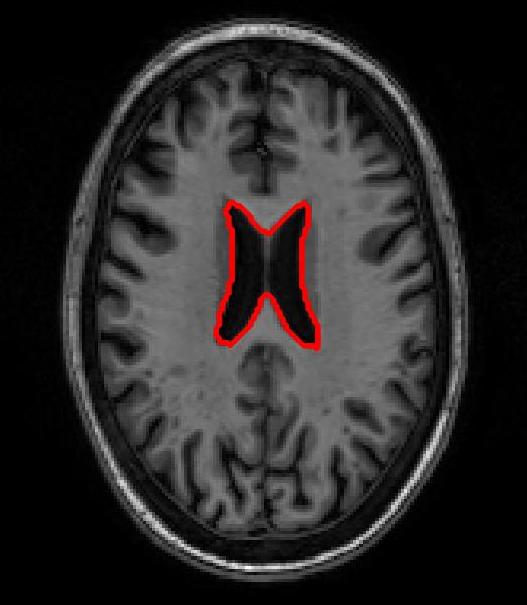} &
\includegraphics[height=2.4cm,width=2.3cm]{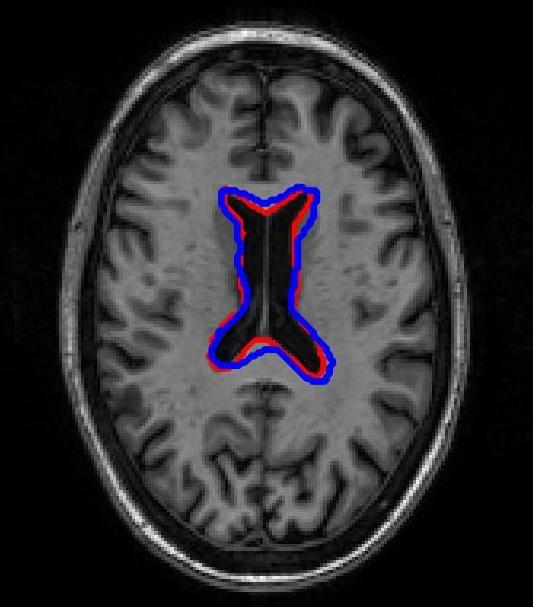} &
\includegraphics[height=2.4cm,width=2.3cm]{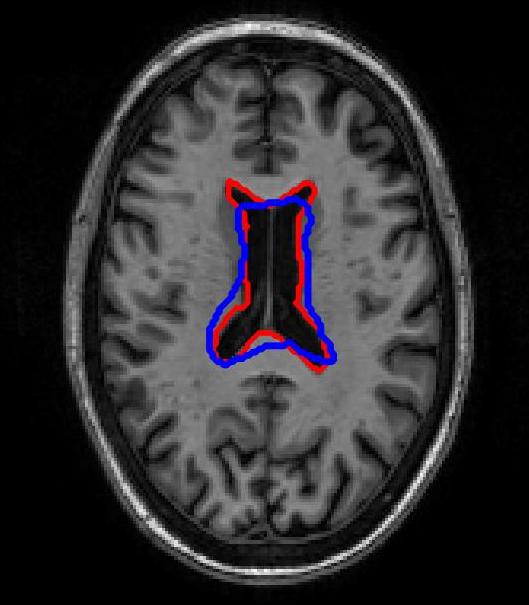} &
\includegraphics[height=2.4cm,width=2.3cm]{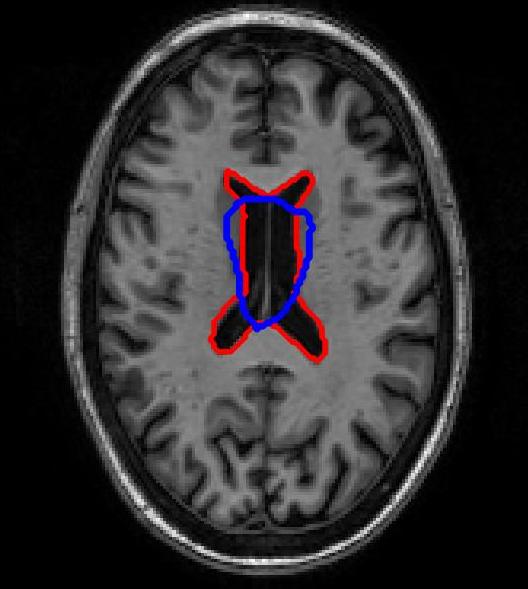} \\
%
(a) & (b) & (c) &(d) & (e) \\
\end{tabular}
\caption{Results for atlas based brain MRI image registration. (a) $I_{R}$ (b) $I_{F}$ with manual segmentation in red. Superimposed registered mask (in blue) obtained using: (c) SR-Net; (d) VoxelMorph; (e) SR-Net$_{wL_{Seg}}$.}
\label{fig:brain}
\end{figure}

Figure~\ref{fig:brain} shows results for brain image registration. We show the reference image in  Figure~\ref{fig:brain} (a) followed by an example floating image in Figure~\ref{fig:brain} (b). The ventricle structure to be aligned is shown in red in both images. Figure~\ref{fig:brain} (c)-(e) shows the deformed  structures obtained by applying the registration field obtained from different methods to the floating image and superimposing these structures on the atlas image. The deformed structures from the floating image are shown in blue. In case of perfect registration the blue and red contours should coincide. In this case SR-Net actually does better than VoxelMorph, while SR-Net$_wL_{Seg}$ does significantly worse due to absence of segmentation information. 
%


\section{Conclusion}
\label{sec:concl}

In this work we have proposed a deep learning based registration method that uses self supervised feature
maps to include segmentation information in the registration framework. Use of self supervised 
segmentation maps enables us to include important structural information in scenarios where manual segmentation'
maps are unavailable, which is the case for majority of datasets. Experimental results show that by using 
self supervised segmentation maps the registration results are close to that obtained using manual segmentation maps.
Hence we conclude that self supervised segmentation maps are an effective way of 
replacing manual segmentation maps and obtaining improved registration. This is applicable for majority of medical image analysis tasks where registration is essential and manual segmentation maps are unavailable.

 \bibliographystyle{splncs04}
 \bibliography{MICCAI2020_RegSeg,MyCitations_Conf,MyCitations_Journ}

\end{document}